\documentclass[conference]{IEEEtran}
\IEEEoverridecommandlockouts
\usepackage{cite}
\usepackage{amsmath,amssymb,amsfonts}
\usepackage{graphicx}
\usepackage{textcomp}
\usepackage{xcolor}
\usepackage{booktabs}
\usepackage{url}
\usepackage{hyperref}
\usepackage{balance}
\usepackage{orcidlink}
\def\BibTeX{{\rm B\kern-.05em{\sc i\kern-.025em b}\kern-.08em
    T\kern-.1667em\lower.7ex\hbox{E}\kern-.125emX}}

\graphicspath{{./figures/}}

\begin{document}

\title{DECICE: AI-Driven Scheduling and Digital Twin Integration for the Cloud-HPC-Edge Compute Continuum}

\author{
\IEEEauthorblockN{
Aasish Kumar Sharma\textsuperscript{1}\orcidlink{0000-0002-7514-2340},
Felix Stein\textsuperscript{1},
Mirac Aydin\textsuperscript{2},
Michael Bidollahkhani\textsuperscript{1}\orcidlink{0000-0001-8122-4441}\\
Sachin P. Nanavati\textsuperscript{3}\orcidlink{0000-0002-5525-4329},
Mohsen Seyedkazemi Ardebili\textsuperscript{4}\orcidlink{0000-0002-1166-6559},
Giorgi Mamulashvili\textsuperscript{1},
Mojtaba Akbari\textsuperscript{1}\\
Jonathan Decker\textsuperscript{1}\orcidlink{0000-0002-7384-7304},
Zoya Masih\textsuperscript{2}\orcidlink{0009-0004-1484-229X},
Julian M. Kunkel\textsuperscript{1,2}\orcidlink{0000-0002-6915-1179}}
\IEEEauthorblockA{%
\textsuperscript{1}Georg-August-Universit\"at G\"ottingen,
\textsuperscript{2}GWDG mbH, G\"ottingen, Germany,
\textsuperscript{3}NAG, Oxford, UK,
\textsuperscript{4}Universit\`a di Bologna, Italy}
}

\maketitle

\begin{abstract}
This paper presents the DECICE project (Device Edge Cloud Intelligent Collaboration framEwork), a Horizon Europe Research and Innovation Action (Grant No.\ 101092582, December 2022 to November 2025) that developed an open-source framework for intelligent workload scheduling across the cloud-HPC-edge compute continuum. A consortium of 12 partners across 6 European countries organized the work into six work packages covering AI-driven scheduling, digital twin infrastructure, system architecture and integration, monitoring, use case validation, and dissemination. The two core technical contributions are an Integrated AI Scheduler (IAIS) employing RNN-based prediction and formal workflow modeling for constraint-aware workload mapping, and a Digital Twin aggregating real-time metrics with carbon intensity and anomaly prediction for energy-aware scheduling. The framework operates within Kubernetes environments, supports unified workflow ingestion from multiple formats, and bridges cloud-native and HPC orchestration through a Slurm integration layer. We present the project vision, the overall architecture, contributions from each work package, quantitative evaluation results, and the open-source release.
\end{abstract}

\begin{IEEEkeywords}
compute continuum, Heterogeneous HPC Systems, AI-Driven Scheduling, Workflow Scheduling, digital twin, Kubernetes, energy-aware computing, Horizon Europe
\end{IEEEkeywords}

\section{Project Overview}

\subsection{Project Details}

DECICE (Device Edge Cloud Intelligent Collaboration framEwork) is a Horizon Europe Research and Innovation Action funded under programme HORIZON-CL4-2022-DATA-01-02, Grant Agreement No.\ 101092582. The project ran from December 2022 to November 2025 and is currently in its dissemination and community adoption phase. The consortium comprised 12 active partners across 6 countries: Georg-August-Universit\"at G\"ottingen (UGOE, coordinator) and GWDG (Germany); E4 Computer Engineering and Consorzio TOP-IX (Italy); KTH Royal Institute of Technology (Sweden); University of Stuttgart / HLRS (USTUTT, Germany); Huawei Technologies D\"usseldorf (HWDU, Germany); SYNYO GmbH (Austria); Marmara University (MARUN, Turkey) and BIGTRI (Turkey); Universit\`a di Bologna (UNIBO, Italy); and the Numerical Algorithms Group (NAG, UK).

\subsection{Motivation and Objectives}

Modern computing workloads span heterogeneous infrastructure ranging from IoT devices and edge nodes through cloud platforms to HPC clusters. This compute continuum introduces scheduling challenges: workloads must be matched to resources satisfying computational, memory, and data locality constraints while meeting energy, latency, and cost requirements. DECICE addressed six objectives: (O1)~leverage the full cloud-HPC-edge continuum; (O2)~build an AI scheduler with dynamic load balancing, energy-efficiency, and green energy awareness; (O3)~design APIs increasing control over network, computing, and data resources; (O4)~implement a Dynamic Digital Twin with AI-based prediction; (O5)~demonstrate usability through real-life use cases; and (O6)~enable trustworthy and security-compliant service deployment.

\subsection{Relevance to COMPSAC Themes}

DECICE addresses COMPSAC symposia themes including Applied Artificial Intelligence (AI-driven scheduling), Autonomous Systems (adaptive workload placement), Computer Architecture and Platforms (Kubernetes federation across HPC, cloud, and edge), Smart IoT Systems (Digital Twin and edge monitoring), and Security, Privacy and Trust (trustworthy deployment under O6).

\section{Work Package Structure and Contributions}

\begin{figure}[t]
    \centering
    \includegraphics[width=\columnwidth]{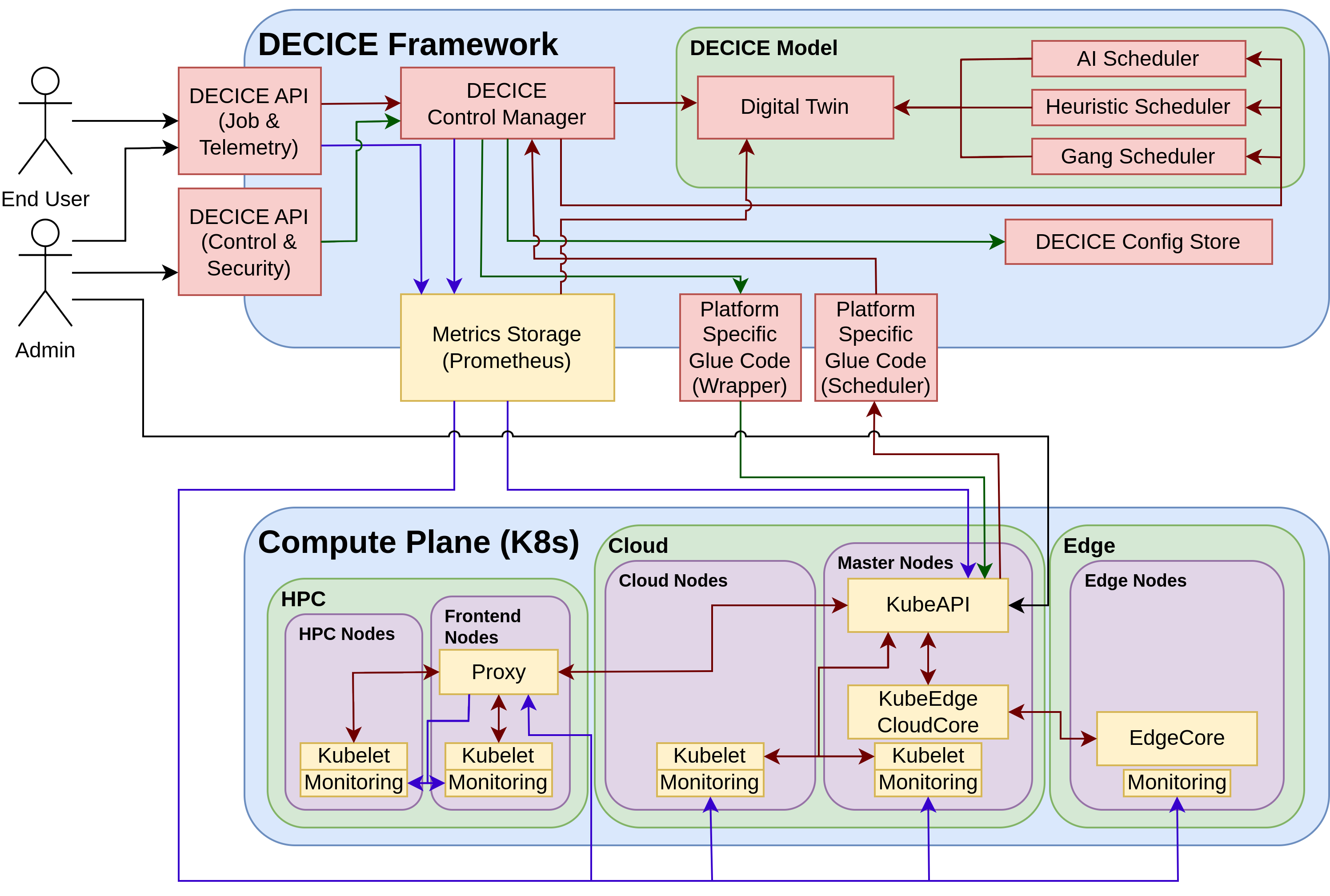}
    \caption{DECICE framework architecture integrating contributions from all work packages. Upper layer: DECICE Framework (WP2 AI Scheduler and Digital Twin, WP3 Control Manager and APIs). Lower layer: Kubernetes-based Compute Plane (WP4 monitoring, WP5 deployment) federating HPC, cloud, and edge resources.}
    \label{fig:arch}
\end{figure}

The project was organized into six work packages. Fig.~\ref{fig:arch} shows the overall framework architecture that integrates contributions from all WPs.

\textbf{WP1: Project Management} (Lead: UGOE) coordinated consortium activities, technical steering, and milestone tracking across all partners and deliverables.

\begin{figure}[t]
    \centering
    \includegraphics[width=\columnwidth]{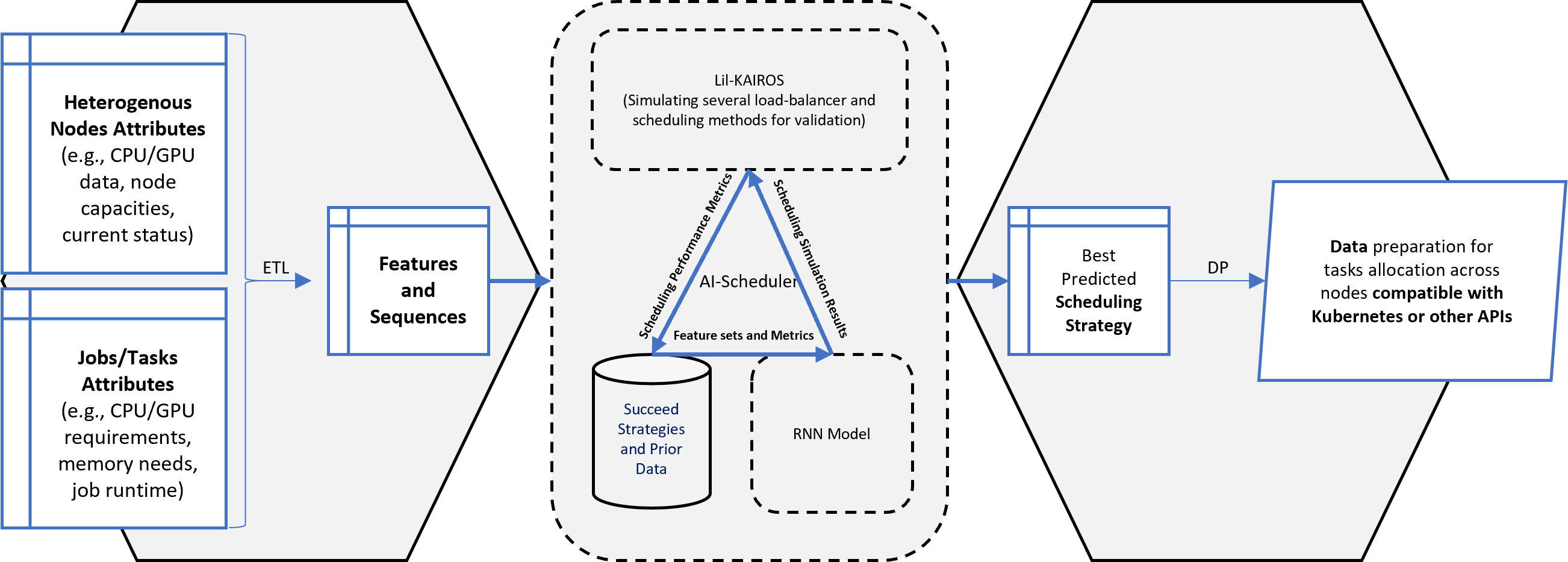}
    \caption{WP2: IAIS data flow. Heterogeneous node and job attributes enter through ETL, pass through the AI scheduler's RNN model and LB-KAIROS simulation, producing Kubernetes-compatible placement decisions.}
    \label{fig:iais}
\end{figure}

\begin{figure}[t]
    \centering
    \includegraphics[width=\columnwidth]{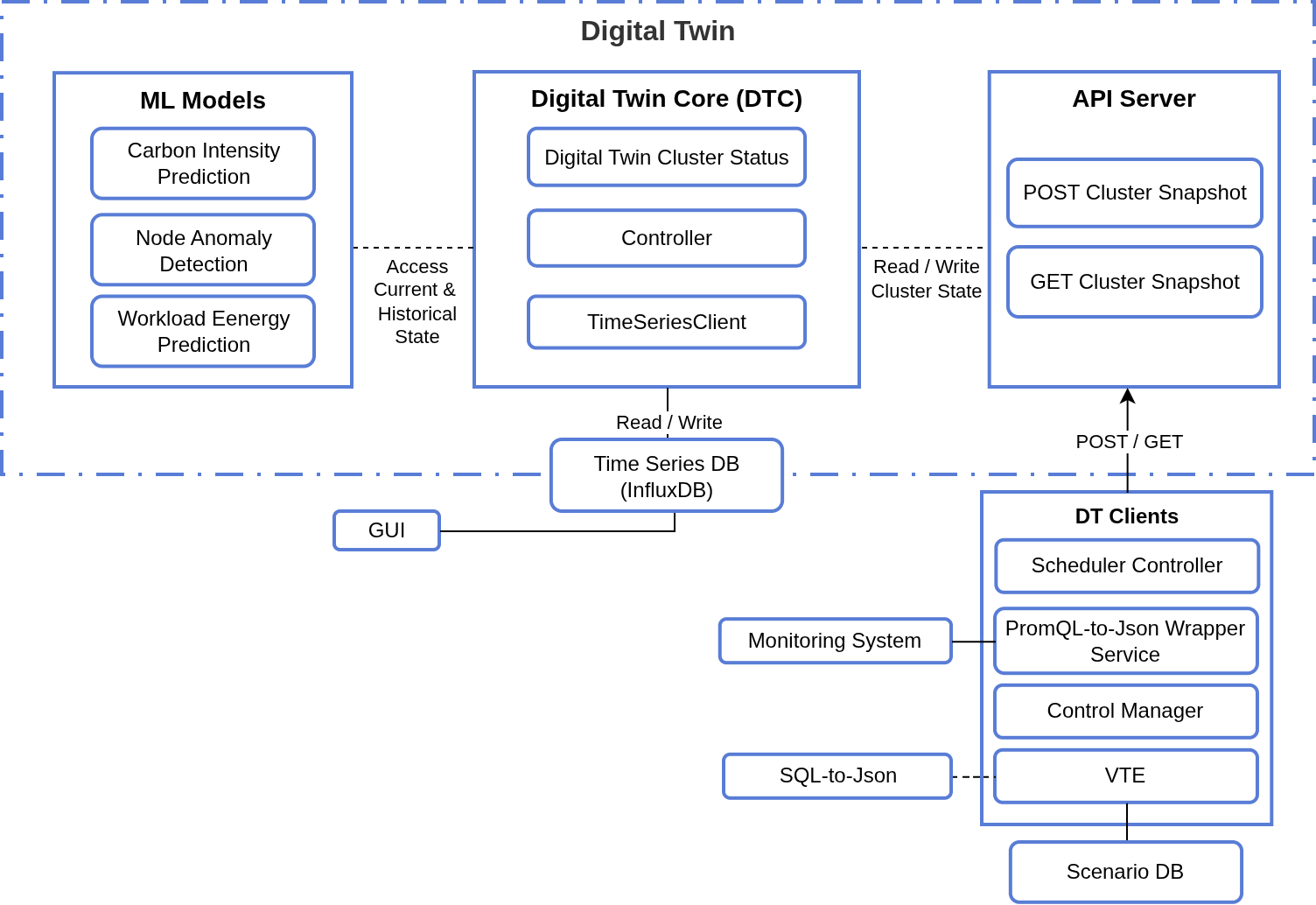}
    \caption{WP2: Digital Twin architecture. The Digital Twin Core (DTC) maintains cluster state via InfluxDB. ML models predict carbon intensity, detect node anomalies, and forecast energy consumption. DT Clients integrate with the scheduler, control manager, and virtual test environment.}
    \label{fig:dt}
\end{figure}

\textbf{WP2: AI Scheduler for Optimization and Adaptation} (Lead: UGOE, after transfer from FBU; key contributors: GWDG, NAG, UNIBO, MARUN, BIGTRI) developed the two core technical components. The Integrated AI Scheduler (IAIS) optimizes job placement across the compute continuum using multiple strategies. HOSHMAND~\cite{bidollahkhani2024hoshmand}, an RNN-based scheduler, learns from historical job-resource state pairs to predict optimal allocations and eliminate redundant scheduling computations. GrapheonRL~\cite{sharma2025grapheonrl}, a Graph Neural Network and Reinforcement Learning framework, models workflows as dependency-aware graphs enabling constraint-aware scheduling without mathematical reformulation. The workflow-driven modeling framework~\cite{sharma2025workflow} formally decomposes heterogeneous HPC scheduling into task-node mapping and schedule derivation, benchmarking exact solvers (MILP via PuLP, SCIP, Gurobi, OR-Tools), constraint programming (CP-SAT), and heuristics (GA, PSO, ACO, SA, HEFT, OLB) across progressively complex workflow scenarios. Fig.~\ref{fig:iais} illustrates the IAIS data flow, and Fig.~\ref{fig:dt} shows the Digital Twin architecture developed by UNIBO, UGOE, and GWDG, which provides node-level power consumption metrics, carbon intensity prediction~\cite{ardebili2024hazardnet}, and anomaly detection~\cite{molan2024graafe} to enable energy-aware scheduling decisions.

\textbf{WP3: Open Framework and Virtual Training Environment} (Lead: GWDG; key contributors: UGOE, E4, USTUTT, HWDU, TOP-IX) designed the overall system architecture, the DECICE Control Manager, and the API layer. WP3 also developed the Synthetic Test Environment for training and validating AI scheduler models, the unified workflow ingestion layer accepting Kubernetes, Snakemake, and Argo Workflows formats, the Kubernetes-Slurm Integration (KSI) framework~\cite{decker2025ksi} enabling Kubernetes workloads on rootless HPC systems, and ephemeral cluster management~\cite{decker2025ephemeral}. The final architecture (D3.2) defines how all components interact as modular, portable services across the federated infrastructure.

\textbf{WP4: Cloud Management Framework Integration} (Lead: USTUTT; key contributors: GWDG, UGOE, KTH, HWDU) implemented the Prometheus-based monitoring stack that collects metrics from HPC, cloud, and edge nodes, including CPU/GPU utilization, memory pressure, network throughput, and storage I/O. The monitoring data feeds both the Digital Twin and the AI Scheduler's decision pipeline. Fault-tolerance mechanisms for Kubernetes and Slurm in HPC environments were also investigated~\cite{aydin2024advcomp}.

\textbf{WP5: Deployment, Validation and Performance Assessment} (Lead: E4; key contributors: UGOE, GWDG, USTUTT, TOP-IX, MARUN, BIGTRI, UNIBO) specified the development environment, deployed the integrated framework across two project phases, and conducted performance evaluation across real-world use cases including cloud robotics (Marmara/BIGTRI), scientific computing (KTH/PDC), and edge inference. The performance evaluation report (D5.5) validated the complete DECICE stack under production-like conditions.

\textbf{WP6: Dissemination and Exploitation} (Lead: SYNYO) managed community engagement, with tutorials at ISC-HPC 2024 and 2025, the open-source release on GitHub, and an exploitation strategy (D6.2) outlining sustainability pathways for the framework beyond the funded period.

\section{Evaluation Results}

\begin{figure}[t]
    \centering
    \includegraphics[width=\columnwidth]{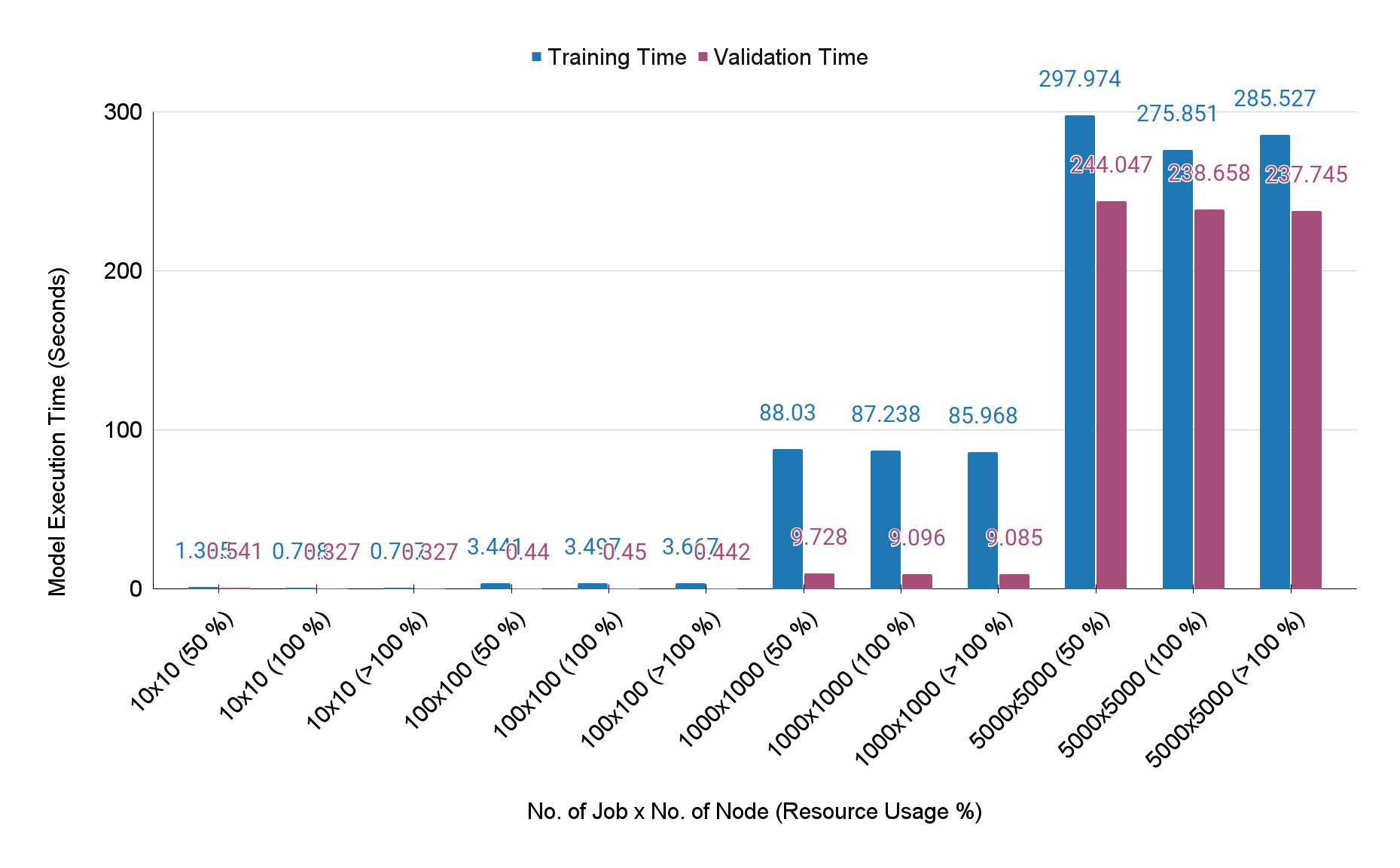}
    \caption{IAIS scalability: training and validation times across problem sizes up to 5000 jobs $\times$ 5000 nodes at 50\% and 100\% resource utilization.}
    \label{fig:scale}
\end{figure}

Fig.~\ref{fig:scale} presents the IAIS scalability evaluation, measuring training and validation times from 10 jobs on 10 nodes to 5000 jobs on 5000 nodes at varying resource utilization levels. Even at the largest scale with 100\% utilization, execution completes within 300 seconds, demonstrating practical applicability to production-sized clusters.

\begin{figure}[t]
    \centering
    \includegraphics[width=\columnwidth]{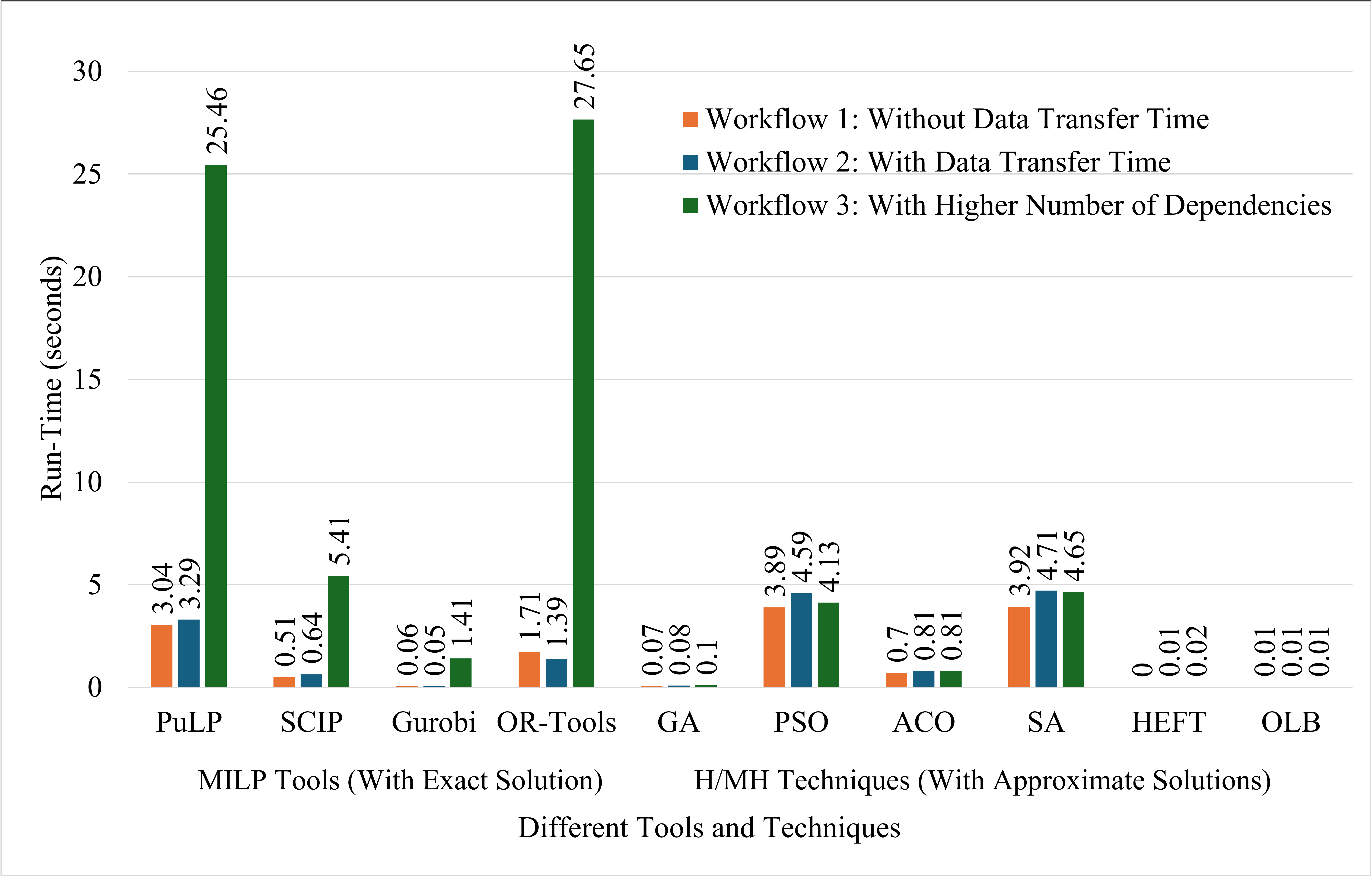}
    \caption{Runtime comparison of 10 scheduling tools across three workflow scenarios. PuLP and OR-Tools scale steeply under high dependency complexity (Workflow~3); Gurobi's internal presolving maintains efficiency; heuristics remain sub-second throughout.}
    \label{fig:runtime}
\end{figure}

Fig.~\ref{fig:runtime} compares 10 scheduling tools across three workflow scenarios of increasing complexity. Exact MILP solvers provide optimal solutions but scale steeply under high dependency counts: PuLP reaches 25.46s and OR-Tools 27.65s on Workflow~3, while Gurobi's internal presolving keeps it at 1.41s. Heuristics (GA, PSO, ACO, SA, HEFT, OLB) maintain sub-second runtimes throughout, with HEFT at 0.01 to 0.02s. This benchmarking~\cite{sharma2025workflow} guides scheduling strategy selection within the IAIS based on workload characteristics and time constraints.

The project produced over 15 peer-reviewed publications consortium-wide, spanning IEEE COMPSAC~\cite{sharma2025workflow, bidollahkhani2024hoshmand, sharma2025grapheonrl}, ACM Computing Frontiers~\cite{kunkel2023decice}, Future Generation Computer Systems~\cite{molan2024graafe, ardebili2024hazardnet}, Journal of Supercomputing~\cite{decker2025ephemeral}, IARIA Intelligent Systems~\cite{decker2025ksi}, ADVCOMP~\cite{aydin2024advcomp}, and Springer FTC~\cite{bidollahkhani2025ftc}. All software components were released under open-source licenses. Tutorials reached over 100 participants at ISC-HPC 2024 and 2025.

\section{Conclusion}

DECICE delivered a validated, open-source framework for AI-driven workload scheduling across the cloud-HPC-edge compute continuum. Through six work packages and 12 consortium partners, the project produced an Integrated AI Scheduler with RNN-based prediction and formal workflow modeling, a Digital Twin with energy-aware and carbon-aware prediction, a unified Kubernetes-based orchestration layer bridging cloud and HPC environments, a comprehensive monitoring stack, and validated use cases in cloud robotics, scientific computing, and edge inference. The modular architecture enables independent adoption of individual components, and the open-source release ensures continued community development beyond the funded period. Future directions include scaling to production workloads with thousands of tasks, extending GrapheonRL~\cite{sharma2025grapheonrl} to multi-objective scheduling across larger compute continuum deployments, and longitudinal energy-aware scheduling using real carbon intensity data.

\section*{Acknowledgment}
{\footnotesize\textit{\copyright~2026~IEEE. Personal use of this material is permitted. Permission from IEEE must be obtained for all other uses. Accepted at the 50th IEEE Computers, Software, and Applications Conference (COMPSAC 2026), Madrid, Spain, July 7--10, 2026.}}

This work was funded by the European Union's Horizon Europe programme, Grant Agreement No.\ 101092582 (DECICE). The authors thank all 12 consortium partners for their contributions across six work packages. Website: \url{https://www.decice.eu}; source code: \url{https://github.com/DECICE-project}.

\balance

\end{document}